\documentstyle[12pt]{article}
\date{\today}

\begin{document}
\title{ Factorizations and Physical Representations}
\author{M. Revzen*,$^{\dag\ddag}$  F.$\;$ C.$\;$ Khanna*,$^{\ddag}$  A.$\;$ 
Mann*,$^{\dag\ddag}$ and$\;$   J.$\;$ Zak*$^{\dag}$\\
$\dag$  Department of Physics, Technion - Israel Institue of Technology,\\
Haifa 32000, Israel\\
$\ddag$  Theoretical Physics Institute,  Department of Physics, University of Alberta,\\
Edmonton,  Alberta, Canada  T6G 2J1.}

\maketitle

\begin{abstract}
A Hilbert space in M dimensions is shown explicitly to accommodate representations that reflect the prime numbers decomposition  of M.
Representations that exhibit the factorization of M into two relatively prime numbers:  the kq representation (J. Zak, Phys. Today, 
{\bf 23} (2), 51 (1970)),
and related representations termed  $q_{1}q_{2}$ representations (together with their conjugates) are analysed, as well as a representation
that exhibits the complete factorization of M. In this latter representation each quantum number varies in a subspace that is associated with
one of the prime numbers that make up M.  

\end{abstract}


\section{Introduction}

Information and computation may be understood in terms of classical 
physics \cite{shannon}. However, the extension of these ideas to the 
quantum
domain \cite{feynman} enriches our understanding of both information theory and quantum mechanics. Thus quantum computers, where entanglement 
and  superposition of states are essential elements, 
 allow computations believed to be intractable on any classical computer. A most often quoted 
example is Shor's \cite{shor} quantum algorithm for factorizing numbers, while there is no known efficient classical algorithm for factoring. 
In this paper
we study the relation of factorizability to quantum physics. Thus we wish to find and characterize physical representations which reflect the prime 
factorization of M, the dimensionality of the space of the problem. Our study is based on Schwinger's \cite{schwinger}  general theory of 
quantum mechanics in finite dimensional space in terms of unitary operators.\\ 
  Schwinger \cite{schwinger}  showed that M-dimensional vector spaces  allow the construction of two unitary operators, U and V (in his
notation), that form a complete operator basis, i.e. they suffice to construct all possible operators of the physical system.
 This
means that if an operator commutes with both U and V it is, necessarily, a multiple of the unit operator. These operators have a period
M, i.e.
\begin{equation} \label{period}
U^{M}\;=\;V^{M}\;=\;1,
\end{equation}
where M is the smallest integer for which this equality holds. The eigenvalues of both U and V  are distinct: they are the M roots of
 unity,  i.e. with $|x\rangle$ the eigenfunctions of U,
$$U|x\rangle\;=\;e^{i({2 \pi \over M})x}|x\rangle,\;\;\;|x+M\rangle\;=\;|x\rangle,\;\;x=1,...,M.$$
The operator V is defined over these eigenvectors as
\begin{equation} \label{step}
V|x\rangle\;=\;|x-1\rangle.
\end{equation}
Schwinger then showed that the absolute value of the overlap between {\it any} eigenfunction of U, $|x\rangle$  and any one of V, 
$|p\rangle,$  is a constant:
\begin{equation}
|\langle p|x\rangle|\;=\;{1 \over \sqrt M}.
\end{equation}
Vector bases with this attribute are referred \cite{berge,wooters} to as conjugate vector bases. It was further noted by Schwinger
 \cite{schwinger} that alternative conjugate vector bases may be constructed. For example, we may let $U\;\rightarrow\;U'\;=\;U^{n}$ 
 for $n\;<\;M$ such that it has no common factor with M. $U'$ has, clearly, the same period and eigenvalues as U. The corresponding 
 $V'$ that satisfies the relevant equation, Eq. (\ref{step}), was shown to be some power of V.\\
Our aim in this paper  is to expand Schwinger's analysis and stress its relation to factorization of M, the dimensionality 
of the space. We choose to consider a specific example of the M-dimensional space, namely M points on a line, i.e., we 
consider discretized and truncated spatial coordinate x and its conjugate momentum p  as our M-dimensional space. This  may be realized by
imposing boundary conditions on the spatial coordinate, x, of the wavefunctions under study, $\psi(x),$ and on their Fourier 
transforms, F(p) (we take $\hbar=1$) \cite{zak2}:
$$\psi(x\;+\;Mc)\;=\;\psi(x),\;\;F(p\;+\;{2\pi \over c})\;=\;F(p).$$
Here M is an integer - it is the dimensionality of the Hilbert space, and we term c the ``quantization length''.  As a consequence 
of the above boundary conditions we have that the value of the spatial coordinate, x,  and the value of the momentum, p, are discrete and finite:
$$x\;=\;sc,\;s\;=\;1,...,M;\;\;p\;=\;{2\pi \over Mc}t,\;t\;=\;1,...,M.$$
In this case we may replace the operators x and p by the unitary operators 
\begin{equation} \label{Tt}
\tau(M)\;=\;e^{i({2\pi \over Mc})x};\;T(c)\;=\;e^{ipc}.
\end{equation}
These operators satisfy the basic commutator relation
\begin{equation}
\tau(M)T(c)\;=\;T(c)\tau(M)e^{-i{2\pi \over M}}.
\end{equation}
They exhibit the dimensionality (i.e. periodicity) automatically (cf. Eq. ({\ref{period})):
\begin{equation}
[\tau(M)]^{M}\;=\;[T(c)]^{M}\;=\;1,
\end{equation}
and we may associate Schwinger's operator $U$ with $\tau(M)$ and his  $V$ with $T(c)$ (henceforth c=1).\\
For our analysis it is convenient to represent the number  M in terms of prime numbers, $P_{j}$, 
\begin{equation} 
M\;=\;\prod_{j=1}^{N}P_{j}^{n_{j}},\;\;P_{j}\;\neq \;P_{i},\;j\;\neq \;i,
\end{equation}
where the $n_{j}$ are integers, and more concisely we denote $P_{j}^{n_{j}}\;{\rm by}\;m_{j},$ i.e.
\begin{equation} \label{primes}
M\;=\;\prod_{j=1}^{N}m_{j}.
\end{equation}
We find thus that the greatest common divisor (gcd) among the 
$m_{j}s$ is 1:
\begin{equation} \label{gcd} 
gcd(m_{j},m_{i})\;=\;1,\;\forall\;j\;\neq \;i,
\end{equation} 
i.e. distinct $m_{i}s$ are relatively prime. 
Our aim is to construct representations that reflect explicitly this factorization of M. In our study of the kq representation 
\cite{zak1,mann,mann1}
the above was used to show that the number of kq representations, $\chi(M),$ having conjugate representations that
can be accommodated in the M dimensional space, is simply related to the number of primes, N,  that appear in M (cf. Eq. (\ref{primes})):
\begin{equation}
\chi(M)\;=\;2^{N-1}.
\end{equation}
It should be noted that the familiar finite dimensional Fourier representation is included in this counting. This is reviewed in section II.
 In section III we consider a novel 
representation, closely related to the kq representation,  that we call $q_{1}q_{2}$ representation \cite{mann1}. Here the relation between 
the number of representations follows much the same reasoning as for the kq representation. In section IV we develop a representation that 
exhibits explicitly  the number of prime numbers that comprise M  (cf. Eq. (\ref{primes})). It is in this section that  
the central point of this 
paper is presented, i.e. we exhibit the inter-relation between the dimensionality of the space 
under investigation  and  representations that reflect its  prime number constituents. For the analysis in this section  we note that 
what was required 
above was less restrictive than having all the involved numbers 
relatively prime, 
i.e.  that among every pair of them  
 Eq. (\ref{gcd})  holds. What is required  is that the numbers are relatively prime numbers $[mod\; M]$. This is 
defined as follows  \cite{mann, mann1,ekert}: two numbers $M_{1},\;M_{2}$ such that their product $M_{1}M_{2}\;=\;M$ are said to be relatively 
prime  $[mod\; M]$ if the equation,
\begin{equation} \label{rp}
tM_{1}\;+\;sM_{2}\;=\;0\;[mod\;M]
\end{equation}
has, for the integers [s, t], only the trivial solution, viz $s\;=\;M_{1},\;t\;=\;M_{2}.$ (Note: from their definition $s=1,..,M_{1}$
 and $t=1,..,M_{2}.)$  This does {\it not} preclude a nontrivial common divisor for $M_{1}\;{\rm and}\;M_{2}.$ This more relaxed requirement
allows representations, presented  in this section, wherein every prime number that makes up the dimensionality,  M,  can be 
associated with a subspace which may be labeled by an appropriate quantum number. In section V we note the relation between the number of 
conjugate kq representations (which is the same as $q_{1}q_{2}\;({\rm or}\;k_{1}k_{2})$ representations) and the number of soutions to the equation 
 $x^{2}\;=\;1\;[mod\;M],$ which is used in number theory to factorize a given integer M into two relatively prime factors.    
The last section, section VI, is devoted to some conclusions and discussion.

\section{The kq representation and factorization}
Schwinger  \cite{schwinger}  noted that U and V, with their powers and products,
 generate $M^{2}$ operators which allow expressing
all operators in terms of them. We shall study space dimensionalities, Ms, which are not prime numbers,   
 i.e. $N\;>\;1$ in Eq. (\ref{primes}). We now briefly review our previous results \cite{mann, mann1} to introduce a somewhat different notation 
that is convenient for our later generalization: Consider bi-partitioning 
the product that represents M (Eq. (\ref{primes})) into two factors,
\begin{equation}
M\;=\;M_{1}M_{2}.
\end{equation}
Here $M_{1}$ incorporates one part of the N factors of Eq. (\ref{primes}) and $M_{2}$ contains the other part. Our way of bi-partitioning 
implies that the two numbers, 
$M_{1}\;{\rm and}\; M_{2},$ are relatively prime, viz. 
$gcd(M_{1},M_{2})\;=\;1$. We now introduce: 
$$L_{1}\;=\;{M\over M_{1}},\;\;L_{2}\;=\;{M \over M_{2}}.$$
 In the case at hand we simply have $L_{1}\;=\;M_{2},\;\;L_{2}\;=\;M_{1},$ however in section IV this definition will prove very useful.  
 $L_{1}\;{\rm and}\;L_{2}$ are also relatively prime $mod\;M,$  
cf. Eq. (\ref{rp}), i.e. the equation
\begin{equation}  
sL_{1}\;+\;t\;L_{2}\;=\;0\;[mod\;M]
\end{equation}
has only the trivial solution for the integers [s, t], viz $s\;=\;M_{1},\;t\;=\;M_{2}.$ 
  This implies that the equation (we take $c=1$),
\begin{equation} \label{ts}
x\;=\;sL_{1}\;+\;tL_{2}\;[mod\;M];\;x=1,..,M;\;s=1,...,M_{1};\;t=1,...,M_{2},
\end{equation}
has a unique solution $x$ for every pair $[s,t],$ with x running over its whole range of M values. We note that, in general,
the pair $[s,t]$ that corresponds to $x=1$ is {\it not} $[s=1,t=1].$ We will now show how to modify Eq. (\ref{ts}) to attain this simpler
relation among the solutions: Let us consider the replacements $s\rightarrow s'N_{1}\;[mod\;M_{1}],\;t\rightarrow t'N_{2}\;[mod\;M_{2}]$ 
with $N_{1}$ relative prime to $L_{2}$
 and $N_{2}$ relative prime to $L_{1}.$ Such replacements retain a unique correspondence $ s\leftrightarrow s'\;[mod\;M_{1}]$ and 
$t\leftrightarrow t'\;[mod\;M_{2}]$ \cite{corresp}. In these new variables Eq. (\ref{ts}) is
\begin{equation}\label{tsts}
x\;=\;s'N_{1}L_{1}\;+\;t'N_{2}L_{2}\;[mod\;M];\;x=1,..,M;\;s'=1,...,M_{1};\;t'=1,...,M_{2}.
\end{equation}
We may now choose the $N_{i}$ to assure that the solution $x=1$ corresponds to the pair $[s'=1,t'=1]$ by solving
$$1\;=\;N_{1}L_{1}\;+\;N_{2}L_{2}\;\;[mod M],$$
i.e. \cite{ekert} 
\begin{equation}\label{inverses} 
N_{2}\;=\;L_{2}^{-1}\;[mod\;M_{2}],\;{\rm and}\;N_{1}\;=\;L_{1}^{-1}\;[mod\;M_{1}].
\end{equation}
Now  Eq. (\ref{ts}) can be rewritten with the solution $x\;=\;1$ corresponding to $s\;=\;t\;=\;1$ as
\begin{equation}
x\;=\;sN_{1}L_{1}\;+\;tN_{2}L_{2} \;\;[mod\;M].
\end{equation}  
An alternative presentation of the above which will be useful in later 
sections is as follows:  Recalling that $L_{1}$ and $L_{2}$
are relatively prime $[mod\;M]$ Eq. (\ref{ts}) may be regarded as the
 solution of a set of two congruences,
\begin{eqnarray}
x\;&=\;s\;[mod\;M_{1}]\nonumber \\
x\;&=\;t\;[mod\;M_{2}].
\end{eqnarray} 
The solution of these is \cite{ekert}
\begin{equation}
x\;=\;sN_{1}L_{1}\;+\;tN_{2}L_{2} \;\;[mod\;M].
\end{equation}
To define a kq representation, we use the two commuting 
operators \cite{zak1,mann}
\begin{equation} \label{operators}
\tau(M_{2})\;=\;e^{i({2 \pi \over M_{2}})x};\;T(N_{1}L_{1})\;=\;e^{ipN_{1}L_{1}}.  
\end{equation}
Since $N_{1}L_{1}\;=\;1\;[mod\;M_{1}],$ the equation $[e^{ipN_{1}L_{1}}]^{M_{1}}\;=\;1$ is a minimal equation (i.e., $M_{1}$ is
the smallest number for which it is satisfied). Therefore the eigenvalues of $T(N_{1}L_{1})$ are $e^{i{2\pi \over M_{1}}k},\;k=1,..M_{1}.$
 (In \cite{mann} we used $e^{ipM_{2}}$ instead of the present $T(N_{1}L_{1});$ these two operators have the same eigenvalues and eigenstates,
but enumerated differently. The advantage of $T(N_{1}L_{1})$ is that it shifts the eigenvalues of $\tau(M_{1})$ by unity whereas $e^{ipM_{2}}$
 shifts them by $M_{2}$.) The common eigenvectors of these operators are given by
\begin{eqnarray}\label{ops}
\tau(M_{2})|k_{1},q_{2}\rangle\;&=\;e^{i{2\pi \over M_{2}}q_{2}}|k_{1},q_{2}\rangle \nonumber \\
T(N_{1}L_{1})|k_{1},q_{2}\rangle\;&=\;e^{i{2\pi \over M_{1}}k_{1}}|k_{1},q_{2}\rangle.
\end{eqnarray}  
They define an M-dimensional kq representation that is associated with the particular factorization of $M\;=\;M_{1}M_{2}.$ 
The indices are always associated with the range of the variable, 
thus, e.g. $q_{2}\;=\;1,..,M_{2}.$    In the following we
 shall omit, unless clarity requires otherwise, the numerical indices of q and k, i.e.
 $q_{2}\;\rightarrow\;q,\; k_{1}\;\rightarrow\;k,$
with similar omission for such  indices which will be introduced later.  It should be noted that in this 
notation operators of different indices
commute as is illustrated in Eq. (\ref{operators}).    
To construct the conjugate vector basis  \cite{mann,wooters} we consider the conjugate  pair of (commuting)
 operators:

\begin{equation} \label{KQa}
\tau(M_{1})\;=\;e^{i({2 \pi \over M_{1}})x},\;\;T(N_{2}L_{2})\;=\;e^{ipN_{2}L_{2}},
\end{equation}
and their eigenfunctions
\begin{eqnarray}\label{evKQ}
\tau(M_{1})|K_{2},Q_{1}\rangle\;&=\;e^{i({2 \pi \over M_{1}})Q_{1}}|K_{2},Q_{1}\rangle;\;\;Q_{1}=1,..,M_{1},\nonumber\\
T(N_{2}L_{2})|K_{2},Q_{1}\rangle\;&=\;e^{i({2 \pi \over M_{2}})K_{2}}|K_{2},Q_{1}\rangle;\;\;K_{2}=1,..,M_{2}.
\end{eqnarray}

The basic commutation relations for our operators are: 
\begin{eqnarray}\label{commut}
T(N_{1}L_{1})\tau(M_{1})\;&=\;\tau(M_{1})T(N_{1}L_{1})e^{i({2\pi\over M_{1}})}, \nonumber \\
T(N_{2}L_{2})\tau(M_{2})\;&=\;\tau(M_{2})T(N_{2}L_{2})e^{i({2\pi\over M_{2}})},
\end{eqnarray}
with all other operators commuting. Hence we have   
\begin{equation}
T(N_{1}L_{1})\tau(M_{1})|k,q\rangle\;=\;e^{i({2\pi \over M_{1}})}e^{i({2\pi \over M_{1}})k}\tau(M_{1})|k,q\rangle,
\end{equation}
indicating that $\tau(M_{1})|k,q\rangle$ is, up to a phase factor, $|k+1,q \rangle.$ In a similar fashion one can show
that $T(N_{1}L_{1})|k,q\rangle$ is, again up to a phase factor, $|k,q-1\rangle.$ 
Now, since k is defined $mod\; M_{1}$ and q  is defined $mod\; M_{2},$ 
 successive application of either (and both)  $\tau(M_{1}),T(N_{2}L_{2})$ on any one vector $|k,q\rangle$  
will generate, {\it uniquely},  all the vectors in the set $|k,q\rangle.$  
 Thus all the states of one set may be generated by the 
operators of the other set \cite{proof}.

Returning to our factorization of M in terms of relative primes, Eq.(\ref{primes}), we find  that only bi-partitionings of the N primes are allowed: 
$P^{n}$ may
{\it not} be split by breaking it up into  two powers, say $P^{n_{1}}$ and $P^{n_{2}},\;\;n=n_{1}+n_{2}$ with one factor in $M_{1}$ 
and the other 
in $M_{2},$ i.e., the bi-partitionings are among the groups of $m_{i}s$  ( Eq. (\ref{primes})).  
Thus the number of 
kq representations  that form a complete operator basis for an M dimensional physical system equals the number of possible 
bi-partitionings of M into products of distinct primes that make M (Eq. (\ref{primes})),  i.e., $2^{N-1}$ \cite{mann}.\\ 

To conclude  this  section we give a new derivation for the overlap $\langle kq|KQ \rangle$:
Recalling our discussion above, we supplement Eq. (\ref{ops}) with 
\begin{equation}
\tau(M_{1})|k,q\rangle\;=\; |k+1,q \rangle\;\;{\rm and}\;\;T(N_{2}L_{2})|k,q\rangle\;=\;|k,q-1\rangle,
\end{equation}
and Eq. (\ref{evKQ}) with
\begin{equation}
\tau(M_{2})|K,Q \rangle\;=\;|K+1,Q\rangle,\;\;{\rm and}\;\;T(N_{1}L_{1}|K,Q\rangle\;=\;|K,Q-1\rangle.
\end{equation}
These are valid up to phase factors, that are, conveniently, chosen  to be null \cite{schwinger}. We now evaluate $\langle kq|A|KQ \rangle,$
 where  $A$ stands for each of the four operators that generate the complete operator basis for the case under study,
$$\tau(M_{1}),\;T(N_{1}L_{1}),\;\tau(M_{2})\;{\rm and}\;T(N_{2}L_{2}).$$ 
This leads to the four  relations
\begin{eqnarray}\label{consistancy}  
e^{i{2\pi \over M_{2}}q}\langle kq|KQ\rangle\;&=\;\langle kq|K+1,Q\rangle, \nonumber \\
e^{i{2\pi \over M_{1}}Q}\langle kq|KQ \rangle\;&=\;\langle k-1,q|K,Q\rangle, \nonumber \\
e^{i{2 \pi \over M_{1}}k}\langle kq|KQ \rangle\;&=\;\langle k,q|K,Q-1\rangle, \nonumber \\
e^{i{2 \pi \over M_{2}}K}\langle k,q|KQ \rangle\;&=\;\langle k,q+1|KQ \rangle.
\end{eqnarray}
 These are solved by
\begin{equation}\label{overlap1}
\langle kq|KQ \rangle \;=\;{e^{i(KqM_{1}-kQM_{2}){ 2\pi \over M}} \over \sqrt M},
\end{equation}
which implies  the conjugacy of the two vector bases \cite{berge,wooters}. 

\section{The $q_{1}q_{2}$  representation}

The choice of the two unitary commuting operators  $\tau(M_{2})$ and  $T(N_{1}L_{1})$  (Eq.(\ref{ops}))  as the ones (corresponding to
Schwinger's  U) that define our vector space basis, i.e. the choice of a  kq  representation to study the system, is optional.
 An alternative choice is the two unitary and commuting 
 operators $\tau(M_{2})$ and  $\tau(M_{1})$  \cite{mann1}. We now discuss 
such a choice - it leads  to the representation  that we choose to call the  $q_{1}q_{2}$  representation, since its labels may be 
considered as 
designating the spatial coordinates. This representation is closely related to the kq representation. It exists only when $M_{1}$ and
 $M_{2}$ are relatively prime, in which case the kq representation has a conjugate KQ representation.
 The common eigenfuctions of $\tau(M_{1})\;{\rm and}\;\tau(M_{2})$ are $|q_{1},q_{2}\rangle.$ Thus, with 
\begin{eqnarray}
\tau(M_{1})\;&=\;e^{i{2 \pi \over M_{1}}x}\;=\;\tau(M)^{M_{2}} \nonumber \\
\tau(M_{2})\;&=\;e^{i{2 \pi \over M_{2}}x}\;=\;\tau(M)^{M_{1}},
\end{eqnarray}
the eigenvector equations  are
\begin{eqnarray} \label{qq'}  
       \tau(M_{1})|q_{1},q_{2}\rangle\;&=\;e^{i{2\pi \over M_{1}}q_{1}}|q_{1},q_{2}\rangle;\;q_{1}\;=\;1,...,M_{1}, \nonumber \\  
\tau(M_{2})|q_{1},q_{2}\rangle\;&=\;e^{i{2 \pi \over M_{2}}q_{2}}|q_{1},q_{2}\rangle;\;q_{2}\;=\;1,....M_{2}.
\end{eqnarray}

These provide an alternative vector basis for the M dimensional space. The complete operator basis includes, in addition, 
 the unitary operators, $$T(N_{1}L_{1})\;\;\; {\rm and}\;\;T(N_{2}L_{2}).$$ The eigenvector equations for these operators are
\begin{eqnarray}\label{kk'}
T(N_{1}L_{1})|k_{2},k_{1} \rangle\;=\;e^{i{ 2\pi \over M_{1}}k_{1}}|k_{2},k_{1} \rangle, \nonumber \\    
T(N_{2}L_{2})|k_{2},k_{1} \rangle\;=\;e^{i{ 2\pi \over M_{2}}k_{2}}|k_{2},k_{1} \rangle.    
\end{eqnarray}
These, too, span the space and form the conjugate vector basis to $|q_{1},q_{2}\rangle.$
 A convenient way to demonstrate this
is by showing that the absolute value of the overlap of any member of one basis with the other is independent of either  vector 
\cite{berge,wooters}.  
We may get the expression for the overlap $\langle q_{1},q_{2}|k_{1},k_{2} \rangle$ in much the same way that we got Eq. (\ref{overlap1}). 
The result is
\begin{equation}
\langle q_{1},q_{2}|k_{1},k_{2}\rangle \;=\;{e^{i(q_{1}k_{1}M_{2}+q_{2}k_{2}M_{1}){2 \pi \over M}} \over \sqrt M},
\end{equation}
assuring that the two vector bases are conjugate.\\
We now obtain the overlap $\langle x|q_{1},q_{2}\rangle$ where   $|x\rangle$ is the eigenvector of $\tau(M)$ with eigenvalue 
$e^{i{2\pi \over M}x}.$
 The method is similar to the one  we used above for the overlap of the vectors belonging
 to conjugate vector bases. 
 Thus, since $\tau(M_{1})\;=\;[\tau(M)]^{M_{2}},$ we have     

\begin{eqnarray} \label{xqQ}  
 \langle x|\tau(M_{1})|q_{1},q_{2}\rangle\;&=\;\langle x|q_{1},q_{2}\rangle e^{i{2\pi \over M_{1}}q_{1}}\;= \nonumber \\
 =\;\langle x|[\tau(M)]^{M_{2}}|q_{1},q_{2}\rangle\;&=\; e^{i{2\pi \over M_{1}}x}\langle x|q_{1},q_{2}\rangle.
\end{eqnarray}
Using  a similar equation with $\tau(M_{2})$ replacing  $\tau(M_{1}),$  we obtain
\begin{eqnarray}\label{ch1}
x\;&=\;q_{1}\;[mode\;M_{1}], \nonumber \\
x\;&=\;q_{2}\;[mode\;M_{2}].
\end{eqnarray} 

Noting that $gcd(M_{1},\;M_{2})\;=\;1$ and using the Chinese Remainder 
Theorem \cite{ekert,zimmer},  we have  that
 the unique solution is 
\begin{equation}
x\;=\;q_{1}N_{1}L_{1}\;+\;q_{2}N_{2}L_{2} [mode\;M] .
\end{equation}
Here $N_{i}\;=\;L_{i}^{-1}\;[mod\;M_{i}],\;\;i\;=\;1,2;$  (cf. Eq.(\ref{inverses})).
Thus we obtain
\begin{equation}
\langle x|q_{1},q_{2}\rangle\;=\;\Delta(x\;-\;q_{1}N_{1}L_{1}\;-\;q_{2}N_{2}L_{2}), 
\end{equation}
with $\Delta(y)\;=\;1$ when $y=0\;[mod\; M],$ and is zero otherwise. The relation for the conjugate vector basis $|k_{1},k_{2}\rangle$ can be 
handled similarly and we get
\begin{equation}
\langle k|k_{1},k_{2}\rangle\;=\;\Delta(k\;-\;k_{1}L_{1}\;-\;k_{2}L_{2}).
\end{equation}  
We now comment briefly on some localization attributes of wave functions when described in this representation. We consider
a state $|\psi\rangle$ that is smeared over one spatial label but is localized in the other:
\begin{equation}
\langle q_{1},q_{2}|\psi \rangle\;=\;{\delta_{q_{1},M_{1}} \over \sqrt {M_{2}}}.
\end{equation}

In the $k_{1}k_{2}$ space we have
\begin{equation}
\langle k_{1}k_{2}|\psi\rangle\;=\;{1 \over \sqrt {MM_{2}}}\Sigma_{q_{2}}e^{2\pi i({q_{2}k_{2}\over M_{2}})}\;=\;{\delta_{k_{2},M_{2}} 
\over \sqrt {M_{1}}}.
\end{equation}
Thus states spread over $q_{2}$ and localized in $q_{1}$ are, in the conjugate basis,  spread in $k_{1}$ and localized in $k_{2},$
with the localization exhibiting the factorization of M.

\section{Complete factorization}     

We now proceed  and obtain a representation in which each prime number in the expression for 
M (cf. Eq. (\ref{primes})) has characteristics of a degree of freedom \cite{schwinger}. We define 
\begin{equation}
L_{j}\;\equiv \;\prod_{k\ne j}P_{k}^{n_{k}}\;=\;{M \over m_{j}};\;\;m_{j}\;=\;P_{j}^{n_{j}}.
\end{equation}
Now consider
\begin{eqnarray}
\tau(m_{j})\;&=\;\tau(M)^{L_{j}}\;=\;U_{j}\;=\;e^{i{2 \pi \over m_{j}}x}, \nonumber \\ 
   T(N_{j}L_{j})\;&=\;T(c)^{N_{j}L_{j}}\;=\;V_{j}\;=\;e^{iN_{j}L_{j}p}.
\end{eqnarray}
We have clearly
\begin{equation}
U_{j}^{m_{j}}\;=\;V_{j}^{m_{j}}\;=\;1,
\end{equation}
which defines the dimensionality of of the relevant coordinates (see below), 
and
\begin{equation}
U_{i}U_{j}\;=\;U_{j}U_{i},\;V_{i}V_{j}\;=\;V_{j}V_{i},\;{\rm and}\;V_{i}U_{j}\;=\;U_{j}V_{i},\;\;\forall \;i\not= j.
\end{equation}
However (cf. \cite{schwinger})
\begin{equation}
V_{i}U_{i}\;=\;U_{i}V_{i}e^{i {2 \pi \over m_{i}}}.
\end{equation}
We define the N-indexed wave function $|q_{1},....q_{N}\rangle$ as the 
eigenfunction of the N (commuting) operators $\tau(m_{j}),\;j=1,...N.$
\begin{equation}
\tau(m_{j})|q_{1},...q_{j},..,q_{N}\rangle\;\equiv\;U_{j}|q_{1},...q_{j},..,q_{N}\rangle\;=\;e^{i{2 \pi \over m_{j}}q_{j}}
|q_{1},...q_{j},..,q_{N}\rangle;\;\;q_{j}=1,...,m_{j}.
\end{equation}
Since the $m_{j}s$ are relatively prime and the equation 
$\tau(m_{j})^{m_{j}}\;=\;1$ is a minimal equation, 
 its  $m_{j}$ eigenfunctions are distinct and different for each index j.
We now relate this wavefunction to the eigenfunction of $\tau(M)$ by the same procedure that we used above: We establish the correspondence
between the M eigenvectors of $\tau(M)$  and those  of $\tau(m_{j}).$ We have N equations of the form 
\begin{eqnarray}\label{Neq}  
     \langle x|e^{i{2\pi \over m_{j}}x}|q_{1},...q_{j},..q_{N}\rangle\;&=\;e^{i{2\pi \over m_{j}}q_{j}}\langle x|q_{1},...q_{j},..q_{N}\rangle\;=
 \nonumber \\
=\;\langle x|[e^{i{2\pi \over M}x}]^{L_{j}}|q_{1},...q_{j},..q_{N}\rangle\;&=\;e^{i{2 \pi \over m_{j}}x}\langle x|q_{1},...q_{j},..q_{N}\rangle.
\end{eqnarray}
Thus we must have
\begin{eqnarray}\label{ch2}
x\;&=\;q_{1}\;[mod\;m_{1}] \nonumber \\
x\;&=\;q_{2}\;[mod\;m_{2}] \nonumber \\
   &......................  \nonumber \\
x\;&=\;q_{N}\;[mod\;m_{N}]. 
\end{eqnarray}
Since $gcd\;(m_{i},m_{j})\;=\;1,\;\;{\rm for\; all}\;\;i\neq j,$  we have by the Chinese Remainder Theorem \cite{ekert,zimmer} that
\begin{equation} \label{ch3}
\langle x|q_{1},...q_{j},..q_{N}\rangle\;=\;\Delta(x\;-\;\Sigma_{j=1}^{N}q_{j}N_{j}L_{j}).
\end{equation}
This associates each of the M values of x with a unique set  of the $q_{j}s.$\\

The M eigenvectors of the commuting operators $T(N_{j}L_{j}),\;j\;=\;1,...,N$ satisfy
\begin{equation}
T(N_{j}L_{j})|k_{1},..,k_{j},..,k_{N}\rangle\;=\;e^{i{2\pi \over m_{j}}k_{j}}|k_{1},..,k_{j},..,k_{N}\rangle,\;k_{j}=1,..,m_{j}.
\end{equation}
By a procedure analogous to the one used above to derive Eq. (\ref{ch3}), we get here 
\begin{equation}
\langle k|k_{1},....k_{N}\rangle\;=\;\Delta(k\;-\;\Sigma_{j=1}^{N}k_{j}L_{j}).
\end{equation}
The overlap is evaluated to be
\begin{equation}
\langle k_{1}...k_{N}|q_{1}...q_{N}\rangle\;=\;{e^{i(\sum_{j=1}^{N}k_{j}q_{j}L_{j}){2\pi \over M}} \over \sqrt M}.
\end{equation}
In the above, the conjugate vector bases representations 
$|q_{1},...q_{j},..q_{N}\rangle\;{\rm and}\;|k_{1},..,k_{j},..,k_{N}\rangle$ exhibit 
the prime  numbers constituents of M. Each index j may be viewed as defining a subspace that is associated with the prime  $P_{j}.$ We refer 
to this representation as the completely factorized representation.\\

\section{Characterization of factorization}

In this section we characterize the possible bi-factorizations of M into two relative primes by the roots of an equation implied by the Chinese
 Remainder Theorem. In principle one might expect that such a process could be reversed, i.e. by noting 
the characteristics of the factorizable physical system, given in some space dimensionality M, one may deduce the factors involved.
However we address ourselves to the former issue. Thus we will show, in parallel with the number theory analysis, that the eigenvalues 
of  unitary operators which form a complete operator basis, \cite{schwinger}, for a given space dimesionality, M, reflect the factors that 
make up the number M.\\

Our analysis above  and, in particular, the completely  factorized  representation as such, allows viewing the N distinct prime
constituents of M, Eq.(\ref{primes}), as N degrees of freedom 
(cf. \cite{schwinger,wooters,ulf}). Now the relation between  $|x\rangle,$
 the eigenfunction of $\tau(M)$  which deals with the space as a 
whole (Eq. (4)), to the eigenfunction $|q_{1},...,q_{N}\rangle$ of 
 $\tau(m_{r}),$ that reflects the  subspaces, each associated with a particular prime $P_{r}$  (and dimensionality $m_{r}$) is given 
by Eq.(\ref{ch3})

$$\langle x|q_{1},....q_{N}\rangle\;=\;\Delta(x\;-\;q_{1}N_{1}L_{1}\;-\;q_{2}N_{2}L_{2}\;-\;.....-\;q_{N}N_{N}L_{N}).$$

As was noted in the previous section this equation brings into our analysis the results of the Chinese Remainder Theorem \cite{ekert,zimmer}.
This theorem  implies the following 
\begin{eqnarray} \label{chinese}      
x\;&=\;1\;\;[mod\;M]\;\Leftrightarrow\;q_{r}\;=\;1\;[mod\;m_{r}],\;{\rm for\;all\;r} \nonumber \\
x^{2}\;&=\;1\;\;[mod\;M]\;\Leftrightarrow\;q^{2}_{r}\;=\;1\;[mod\;m_{r}],\;{\rm for\;all\;r} .
\end{eqnarray}
The  equation $x^{2}\;=\;1\;[mod\;M]$ has several solutions. We will henceforth designate the solutions by $a_{s}.$ We have immediately
that, if  $a_{s}$ is a solution, viz $a_{s}^{2}\;=\;1 \;[mod\;M],$ so is $-a_{s},$ i.e the solutions appear in pairs.\\
We now argue that the number of pairs of solutions is $2^{N-1}.$  Thus we may associate each solution with a conjugate
pair of the kq-representation (or equivalently with the  $q_{1}q_{2}$  and $k_{1}k_{2}$ representations) that can be 
accommodated in  M dimensions. 
The trivial solution,
 $a_{s}\;=\;1,$ is always (i.e. even if M is (power of) prime) present. It corresponds to the trivial factorization,
 $M\;=\;1\cdot M$ that we associate with the Fourier representation \cite{mann,mann1}. We now show that the number of solutions to 
$x^{2}\;=\;1\;\;[mod\;M]$ equals $2^{N-1}.$ The proof is direct:  Eq.(\ref{chinese})  implies that 
$$x^{2}\;=\;1\;[mod\;M]\;\Rightarrow \; q_{r}\;=\;\pm\;1 \;[mod\;m_{r}]\;{\rm for\; r}\;=\;1,...,N.$$
This gives $2^{N}$ possibilities. But only half of these are distinct since the two solutions $a_{s}\;=\;\pm\;1$ give equivalent 
factorization but in a reverse order  (if $a_{s}$ satisfies $(a_{s}\;+\;1)(a_{s}\;-\;1)\;=\;0 \;[mod\;M],$ then $-a_{s}$ satisfies
$(a_{s}\;-\;1)(a_{s}\;+\;1)\;=\;0\;[mod\;M]),$ and as the order of the factors is immaterial the two lead to one bi-factorization. Note that
similar reasoning introduces a factor $1/2$ in counting the number of kq-representations; there this was interpreted as having each
distinct bi-factorization leading to a distinct conjugate pair of vector bases - the kq and KQ  \cite{mann}. Thus $2^{N-1}$ gives
the number of kq conjugate pairs and the number of solutions of $x^{2}=1\;[mod\;M]$, both expressing the bi-factorization of M into 
coprime numbers.\\
To clarify the above we now consider, in some detail, a simple example: Let $M\;=\;105\;=\;3\cdot5\cdot7.$ Thus we have 
\begin{eqnarray}\label{character1}
m_{1}\;&=\;3,\;N_{1}\;=\;2,\;L_{1}\;=\;35; \nonumber \\
m_{2}\;&=\;5,\;N_{2}\;=\;1,\;L_{2}\;=\;21; \nonumber \\   
m_{3}\;&=\;7,\;N_{3}\;=\;1,\;L_{3}\;=\;15.
\end{eqnarray}
There are $2^{2}\;=\;4$ pairs of (distinct) solutions
\begin{eqnarray}\label{character2}
q_{1}\;&=\;q_{2}\;=\;q_{3}\;=\;1,\;\Rightarrow\;a_{1}\;=\;1\;\; [mod\; 105], \nonumber \\
q_{1}\;&=\;q_{2}\;=\;1,\;q_{3}\;=\;-1,\;\Rightarrow\;a_{2}\;=\;76\;[mod\;105], \nonumber \\
q_{1}\;&=\;1,\;q_{2}\;=\;q_{3}\;=\;-1,\;\Rightarrow\;a_{3}\;=\;34\;[mod\;105], \nonumber \\
q_{1}\;&=\;q_{3}\;=\;1, q_{2}\;=\;-1,\;\Rightarrow\;a_{4}\;=\;64\;[mod\;105].
\end{eqnarray}
The four other solutions may be obtained by reversing the signs of the $a_{s}$ which is obtained  by changing the signs of all three $q_{r}$ in 
each set. One can readily check that $a_{s}^{2}\;=\;1\;[mod\;105]$ in all cases. Now we  have it that, for each s $(s\;=\;2,3,4)$
$$(a_{s}\;+\;1)(a_{s}-1)\;=\;0\;\;[mod\;105].$$
Inserting the values of the $a_{s}$ from $s\; =\; 2$ to $s\; =\; 4$ (skipping the trivial case of $s\;=\;1$)
we get the following expressions   for $(a_{s}\;+\;1)(a_{s}-1)$
\begin{eqnarray}
 s\;&=\; 2:\;\; 5\cdot11(15)(7),\nonumber \\
 s\;&=\; 3:\;\;11(15)(7), \nonumber \\
 s\;&=\; 4:\;\;3\cdot13(21)(5),
\end{eqnarray}
all evidently  zero  $[mod\; 105]$.
We see that every distinct root leads to a distinct bi-factorization. Since the bi-factors must be distinct in every case, so must be
the $a_{s}.$\\
To summarize, we have shown that among the eigenstates of the 
completely factorized representation, those distinguished by 
$q_{j}\;=\;\pm1\;(j=1,...N)$ correspond uniquely to the relatively prime 
bi-factorization of M.\\

\section {Conclusions and discussion}
Shor's discovery \cite{shor} of an algorithm for factorization with quantum computers forms a central step in the development
of quantum information theory. The number theoretic basis of the factorization  method in Shor's algorithm has been  studied extensively 
 \cite{ekert}. In this paper we give  what may be viewed  as  a  study of the physics of factorization, i.e.
the inter-relation between the dimensionality of the space under investigation and the representations that reflect its prime number constituents.
 To this end we elaborate on Schwinger's \cite{schwinger} analysis of unitary operator bases for finite 
dimensional Hilbert spaces and show, in what we consider to be a  physical language, that a natural representation is available which exhibits
the prime number constituents of M. In such a representation each of the N prime numbers present in the prime factorization 
of M defines a subspace. We give the operator basis acting in such subspaces. We further show that different, when possible, bi-factorizations 
of M may 
be viewed as different conjugate pairs of vector bases that may be associated with the kq representations \cite{zak2, zak1}, or $q_{1}q_{2}$
  and $k_{1}k_{2}$
representations. It was 
shown that the factorization of the dimensionality of the space as a number is equivalent to the breakup of the space into subspaces each 
representing a distinct degree of freedom that reflects a prime number that is among the prime constituents of M.\\   

\bigskip
{\bf Acknowledgements}\\
FCK acknowleges the support of NSERC. AM and MR thank The Theoretical Physics Institute for partial support, and the National
University of Singapore and in particular Professor B. -G. Englert for kind hospitality and helpful discussions. \\

\bigskip
$ \ast $ {\bf electronic addresses:} revzen@physics.technion.ac.il,  khanna@phys.ualberta.ca,\\  
ady@physics.technion.ac.il,  zak@physics.technion.ac.il.


\begin{thebibliography}{99}
\bibitem{corresp} An illustrative example is: $L_{1}=5,\;L_{2}=3;\;N_{1}=2,\;N_{2}=2$ which leads to the correspondences
$s=1,2,3\;\leftrightarrow\;s'=2,1,3;\;t=1,2,3,4,5\;\leftrightarrow\;t'=3,1,4,2,5.$
\end{thebibliography}
\end{document}